\documentclass[twocolumn,floatfix,showpacs,aps,nofootinbib]{revtex4}

\usepackage{amssymb,amsmath}
\usepackage{graphicx,float}
\usepackage{indentfirst}
\newcommand{\be}{\begin{equation}}
\newcommand{\ee}{\end{equation}}

\newcommand{\ba}{\begin{eqnarray}}
\newcommand{\ea}{\end{eqnarray}}

\begin{document}

\title{Asymmetrical bloch branes and the hierarchy problem}
\author{A. de Souza Dutra}
\email{dutra@feg.unesp.br}
\author{G. P. de Brito}
\email{gustavopazzini@gmail.com}
\author{J. M. Hoff da Silva}
\email{hoff@feg.unesp.br}
\affiliation{Departamento de F\1sica e Qu\1mica, Universidade
Estadual Paulista, Av. Dr. Ariberto Pereira da Cunha, 333,
Guaratinguet\'a, SP, Brazil.}


\begin{abstract}
We investigate a two scalar fields split braneworld model which leads to a possible approach to the hierarchy problem within the thick brane scenario. The model exhibits a resulting asymmetric warp factor suitable for this purpose. The solution is obtained by means of the orbit equation approach for a specific value of one of the parameters. Besides, we analyze the model qualitative behaviour for arbitrary parameters by inspecting the underlying dynamical system defined by the equations which give rise to the braneworld model. We finalize commenting on the metric fluctuation and stability issues.
\end{abstract}

\maketitle

\flushbottom

\section{Introduction}

The very first motivation for the introduction of braneworld models constructed upon warped spacetimes was the solution of the hierarchy problem \cite{RS1}. Since then, several models were proposed generalizing it in a broad range of aspects. The set up of the model presented in \cite{RS1} comprises two infinitely thin branes placed at the extremes of an $\mathbb{Z}_2$ symmetric orbifold. By simple inspection of the Higgs mechanism in the so called visible brane it is possible to see that the presence of the warp factor, appearing as a ``conformal'' gravitational weight, is quite enough to reduce the energy scales from a fundamental (Planckian) to the TeV (Standard Model) one.

However, in the above scenario the Standard Model fields are assumed, a priori, living on the visible brane. It has at least one problematic consequence. In the usual construction of the particle physics, if it is not imposed any additional symmetry precluding the leptonic (baryonic) number variation, questions as the proton decay shall be faced. Usually this type of problem is circumvented by allowing the existence of a $QQQL$-like operator, but endowed with a very small coupling constant in such a way that its possible effects are only scrutinized at very high energies. Nevertheless, as mentioned, by means of the warp factor, high energies are brought back to TeV.

It was realized in the reference \cite{NIMA}, within the context of non warped spacetimes, that this problem could be solved if the brane has some thickness. The heuristic argument goes as follows: let the five-dimensional action for the $QQQL$ operator be
\begin{equation}
S\sim \int d^5x \sqrt{-G}\hspace{.2cm} (QQQL) .\nonumber
\end{equation} Supposing the fields living on the brane, it was found that $Q\sim e^{-\rho r^2}q(x^\mu)$ and $L\sim e^{-\rho^2(r-\Delta)^2}l(x^\mu)$ where $r$ is the extra dimension, $\rho$ is a parameter of order of the four-dimensional fermionic mass scale and $\Delta$ accounts for to locate the quarks and fermions wave functions at different places within the brane. It is, hence, the $\Delta$ parameter which brings information about the brane thickness in this example. After to separate out the integral, taking into account a simple gaussian warp factor $e^{-2\tau^2r^2}$ (where $\tau$ is a parameter with units of $[length]^{-1}$, usually related to the inverse of the AdS radius, just for the purposes of this discussion) in order to mimic a thick brane scenario, we arrive at
\begin{eqnarray}
S\sim \exp\!\Bigg(\!\!\!\!-\!\Delta^2\rho^2\Bigg[1-\frac{\rho^2}{\tau^2+\rho^2}\Bigg]\!\Bigg)\sqrt{\frac{\pi}{\tau^2+\rho^2}}\int d^4x (qqql). \nonumber
\end{eqnarray} Therefore the existence of some thickness is responsible for attenuate the effects of the quantum operator, acting as a small effective coupling constant.

The idea of universal extra dimension \cite{UNI}, which all the standard model fields and gravity may experience, has spread over a wide range of thick braneworld models \cite{VAR}. An important point to be stressed is that in the realization of a thick braneworld by most models, a well behaved scalar field perform the domain wall scenario and the brane is understood as located in the maximum slope regium of this scalar field along the extra dimension \cite{GREMM}. Obviously, the solution of the full coupled equations (solitonic scalar field plus gravity) is not an easy task and the use of advanced techniques, based in the supersymmetric quantum mechanics tools (just as the superpotential), are used. In this programme, however, the emergent potentials are quite symmetric (see \cite{FOLO} for an up to date review) around the brane, even in the cases of split branes \cite{SPLIT}. It turns out that the resulting warp factors inherit this symmetry and any attempt to use them to rescale the masses, solving the hierarchy problem, is unsuccessful in these models. Inspite of the importance, there are only a few models looking at this point. A prominent approach \cite{HER} considers the existence of an infinitely thin brane placed in the background generated by a thick braneworld at a finite distance of the scalar field domain wall. In this vein, the hierarchy problem can be addressed once again. By the reasons exposed in the previous paragraphs, this approach sounds unsatisfying. Another work addressing the hierarchy problem within the context of thick braneworlds is found in Ref. \cite{PAPERJHEP}, in which the authors show the possibility of extending the braneworld sum rules and achieve a well behaved scenario with positive brane tension only. In \cite{PAPERJHEP} the authors consider one scalar field in the bulk, but the assume a priori a two-kink form of such a field.

In this paper we present a split brane model which is capable to generate a non-symmetric potential, leading then to the possibility of a well behaved mass rescaling on one of the branes. As it will be shown, we arrive at a two-kink profile for one of the scalar fields. The price to be paid is the necessary inclusion of an extra scalar field in the bulk. The model that will be considered in this work is a generalization of the so-called BNRT model \cite{BNRT}. It is remarkable that the BNRT model was already considered in the context of thick brane solutions \cite{BLOCH}, however, despite of its interesting features, it can not be used to address the hierarchy problem. We will show that the addition of a new term in its superpotential induces the symmetry break on the warp factor, and hence, the hierarchy problem can be addressed. The paper is organized as follows: in the next section, we present the basic model setup studing its qualitative behaviour for arbitrary parameters. Going further, we investigate a particular case in which we show the explicit solutions and analyse its physical consequences. After studying the model stability in Section III, we conclude.

\section{The model}

We consider the action describing five-dimensional gravity coupled to two interacting real scalar fields, namely
\begin{equation}
 S = \int d^4x dr \sqrt{-G} \bigg[ \frac{1}{4} R + \mathcal{L}_{scalar} \bigg],
\end{equation}
where
\begin{equation}
 \mathcal{L}_{scalar} = - \frac{1}{2} \partial_{a}\phi \partial^{a}\phi - \frac{1}{2} \partial_{a}\chi \partial^{a}\chi - V(\phi,\chi).
\end{equation}
We are working in warped space-time, in which the invariant interval may be written as
\begin{equation}
 ds^2 = G_{ab}dx^a dx^b = e^{2A(r)}\eta_{\mu \nu} dx^{\mu}dx^{\nu} + dr^2,
\end{equation}
where $r$ is the extra dimension, $\eta_{\mu \nu}$ is the usual Minkowski metric with $diag(-,+,+,+)$ and $e^{2A(r)}$ is the so-called warp factor. We assumed that the warp factor and the scalar fields only depend on the extra dimension $r$. The resulting equations of motion for this system are
\begin{eqnarray}\label{secondorder1}
 \frac{d^2 \phi}{dr^2} + 4 \frac{dA}{dr} \frac{d\phi}{dr} = \frac{\partial V}{\partial \phi} \quad,\quad \frac{d^2 \chi}{dr^2} + 4 \frac{dA}{dr} \frac{d\chi}{dr} = \frac{\partial V}{\partial \chi}
\end{eqnarray}
and
\begin{eqnarray}\label{secondorder2}
\frac{d^2 A}{dr^2} = - \frac{2}{3}\bigg[\bigg(\frac{d\phi}{dr}\bigg)^2 + \bigg(\frac{d\chi}{dr}\bigg)^2 \bigg],
\end{eqnarray}
\begin{eqnarray}\label{secondorder3}
\bigg(\frac{dA}{dr} \bigg)^2 = \frac{1}{6}\bigg[\bigg(\frac{d\phi}{dr}\bigg)^2 + \bigg(\frac{d\chi}{dr}\bigg)^2 \bigg] - \frac{1}{3} V(\phi,\chi) .
\end{eqnarray}
In general, the above system of differential equations is very complicated, however, it can be simplified if we introduce a superpotential function $W(\phi,\chi)$ that satisfies
\begin{equation}\label{firstorder}
\frac{d \phi}{dr} = W_{\phi} \quad,\quad \frac{d \chi}{dr} = W_{\chi} \quad,\quad \frac{dA}{dr} = -\frac{2}{3}W,
\end{equation}
where $W_{\phi_i}$ stands for the derivative with respect to $\phi_i$. It can be demonstrated that these first-order equations share the same solutions with (\ref{secondorder1}), (\ref{secondorder2}) and (\ref{secondorder3}) since the superpotential function is related with $V(\phi,\chi)$ as follows
\begin{equation}\label{pot}
V(\phi,\chi) = \frac{1}{2} W^2_{\phi} + \frac{1}{2} W^2_{\chi} - \frac{4}{3} W^2.
\end{equation}
Combining the first pair of equations in (\ref{firstorder}) we get the so-called orbit equation
 \begin{equation}\label{orbit}
\frac{d \phi}{d \chi} = \frac{W_{\phi}}{W_{\chi}}.
\end{equation}
By solving this equation are able to decouple the first pair of equations in (\ref{firstorder}), and then get the solutions for the interacting scalar fields. Once we know the solutions for the scalar fields, it becomes a simple task of integration to solve the last equation in (\ref{firstorder}), and as a consequence, determine the warp factor.

Here we propose a model described by the following superpotential
\begin{equation}\label{superpotential}
W(\phi,\chi) = v^2 \phi - \frac{1}{3} \phi^3 - \lambda \phi \chi^2 - \frac{\beta}{3} \chi^3,
\end{equation}
where $\lambda$, $\beta$ and $v$ are real parameters, and we restricted $v$ to be positive. It is interesting to note that the above model represents an extended version of the so-called BNRT model \cite{BNRT}, which can be recovered in the limit $\beta \rightarrow 0$.  Hence, in view of Eq. (\ref{pot}) along with Eq. (\ref{superpotential}) we have the following potential:
\begin{eqnarray}
& &V(\phi,\chi)= -\frac{4}{27}\phi^6 - \frac{4\beta^2}{27}\chi^6 - \frac{4\lambda^2}{3}\phi^2 \chi^4 - \frac{8\lambda}{9} \phi^4 \chi^2 - \nonumber \\ & & - \frac{8\beta}{27}\phi^3\chi^3 - \frac{8\lambda\beta}{9}\phi\chi^5 + \left(\frac{1}{2} + \frac{8v^2}{9}\right)\phi^4 + \nonumber \\ & & + \left(1 + \frac{8v^2}{9} + 2 \lambda \right)\lambda \phi^2 \chi^2 + \left(2\lambda + \frac{4v^2}{9}\right) \beta \phi \chi^3 - \nonumber \\ & & - \left(1 + \frac{4v^2}{3} \right)v^2 \phi^2 - \lambda v^2 \chi^2 - \frac{v^4}{2} .
\end{eqnarray}
This is a rather unusual, though polynomial, potential. Perhaps the best way to justify its consideration is a posteriori, by looking at its physical implications. As we shall see, the aforementioned potential leads to a possible solution to the hierarchy problem within the thick brane scenario. Hence, the considered potential seems to be a worsening of the situation, but its investigation via the, quite simple and useful, orbit equation procedure leads to an important achievement.

\subsection{Dynamical system analysis}

From the equations
\begin{eqnarray}\label{DS}
\frac{d\phi}{dr}=W_{\phi} \quad , \quad \frac{d\chi}{dr}=W_{\chi}\label{DS2}
\end{eqnarray}
along with Eq. (\ref{superpotential}), we have the autonomous dynamical system defined by
\begin{eqnarray}
\frac{d\phi}{dr}\equiv f_1(\phi,\chi)= v^2 -  \phi^2 - \lambda \chi^2,\label{DS3}\\ \frac{d\chi}{dr}\equiv f_2(\phi,\chi)=-2\lambda \phi \chi-\beta \chi^2,\label{DS4}
\end{eqnarray}
whose set $(FP)$ of fixed points $(f_1(\phi,\chi)=0=(f_2(\phi,\chi))$ is
\begin{eqnarray}
FP&=&\left.\Bigg\{(\phi, \chi)/(+v,0),(-v,0),\Bigg(+\frac{v \beta}{\sqrt{\beta^2+4\lambda^3}},\right.\nonumber\\&-&\left.\frac{2 v \lambda}{\sqrt{\beta^2+4\lambda^3}}\Bigg),\Bigg(-\frac{v \beta}{\sqrt{\beta^2+4\lambda^3}},+\frac{2 v \lambda}{\sqrt{\beta^2+4\lambda^3}}\Bigg)\Bigg\}.\right.\nonumber
\end{eqnarray}

In the next Section, we shall present a complete analysis for the particular $\lambda=1$ case. In fact, for this case the orbit equation approach will be quite powerful, allowing for the analytic integration of the system. It is interesting, however, for completeness to investigate some system qualitative properties by exploring the underlying dynamical system for the arbitrary $\lambda$ situation.

We stress that the obtained set of fixed points reduces to the vacuum points of the next Section for $\lambda=1$. Now, following the usual dynamical system theory \cite{DS}, it is necessary to classify the fixed points acording to the sign of the Jacobian matrix eigenvalues real parts. Therefore, the relevant determinant is given by
\begin{eqnarray}
\Bigg|\begin{matrix} -2\phi-\Delta & -2\lambda\chi \\ -2\lambda\chi & -2\lambda\phi-2\beta\chi-\Delta \end{matrix} \Bigg|=0,\nonumber
\end{eqnarray} where $\Delta$ is the eingenvalue. The solution may be easily found, reading
\begin{equation}
\Delta\!=\!\!-[\phi(1+\lambda)+\beta \chi]\!\pm\!\!\sqrt{[\phi(\lambda-1)+\beta\chi]^2+(2\lambda\chi)^2}.\label{DS5}
\end{equation}

By substituting the fixed point $(\phi, \chi)=(+v,0)$ into Eq. (\ref{DS5}), one sees that
\begin{eqnarray}
\Delta_+=-2 v \quad , \quad \Delta_-=-2 v \lambda,\nonumber
\end{eqnarray} while for the case $(\phi,\chi)=(-v,0)$ we have
\begin{eqnarray}
\Delta_+=2 v \quad ,\quad \Delta_-=2 v \lambda.\nonumber
\end{eqnarray} Therefore, the analysis of these fixed points for an arbitrary $\lambda$ is quite easy. For the $(+v,0)$ case, we have a saddle point for $\lambda<0$ and an attractor for $\lambda>0$. For the $(-v,0)$ case, there is the opposite situation, i. e., a saddle point if $\lambda<0$ and a repellor when $\lambda>0$. The study of the other fixed points is somewhat more involved and not particularly elucidative. It is worthwhile, however, to mention an interesting case. Denoting $\varepsilon=\pm 1$, one has the following eigenvalue
\begin{equation}
\Delta=\frac{v}{\sqrt{\beta^2+4\lambda^3}}\Bigg\{\varepsilon\beta(\lambda-1)\pm [\beta^2(\lambda + 1)^2+16\lambda^4]^{1/2} \Bigg\}.\nonumber
\end{equation} Within this notation the fixed points are $\Bigg(\frac{\varepsilon v\beta}{\sqrt{\beta^2+4\lambda^3}},-\frac{2\varepsilon v \lambda}{\sqrt{\beta^2+4\lambda^3}}\Bigg)$, hence for $\varepsilon=+1$ we have the third fixed point of the set $FP$, and for $\varepsilon=-1$ we have the fourth point of $FP$. The region in evidence in Fig. 1 shows the combination of $\beta$ and $\lambda$, for which $\Delta_+<0$ in the $\varepsilon=-1$ case. Thus, for a positive $\lambda$ inside the aforementioned region we may also have $\Delta_-<0$ at the same time, showing the existence of the non-trivial attractor point $\Bigg(-\frac{v \beta}{\sqrt{\beta^2+4\lambda^3}},+\frac{2 v \lambda}{\sqrt{\beta^2+4\lambda^3}}\Bigg)$.

\begin{figure}[H]
\center
\includegraphics[width=10cm]{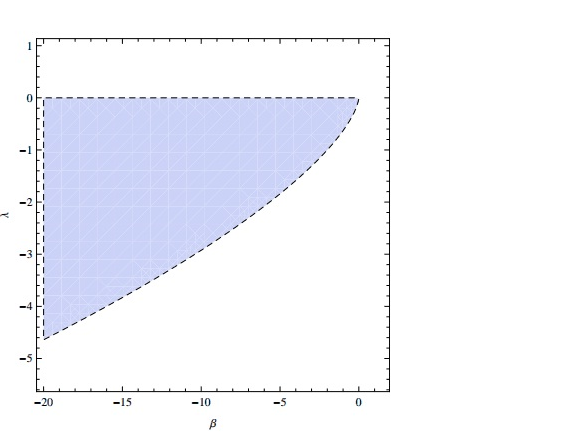}
\label{SUB}
\caption{Interesting region for the existence of an attractor point.}
\end{figure}

\subsection{The $\lambda=1$ case}

Henceforth, it will be considered only the case $\lambda = 1$, for which we arrive at an analytical solution. Using equation (\ref{orbit}) we get
\begin{equation} \label{ORBIT}
\frac{d \phi}{d \chi} = \frac{\phi^2 - v^2 + \chi^2}{2 \phi \chi + \beta \chi^2}.
\end{equation}
Despite of the nonlinearity in the equation above it can be solved analytically. In fact, one can multiply both sides by the integrating factor $\chi^{-2}$ and use the exact differential method to get the implicit solution
\begin{equation}\label{orbit2}
\phi^2 - v^2 + \beta \chi \phi = \chi^2 + c_0 \chi.
\end{equation} 
In fact, eq. (\ref{ORBIT}) may be recast in the following form
\begin{eqnarray}
(2\phi\chi+\beta\chi^2)\frac{d\phi}{d\chi}=\chi^2+\phi^2-v^2,\nonumber
\end{eqnarray} which after the multiplication of the integrating factor $\chi^{-2}$ leads to
\begin{eqnarray}
\frac{2\phi}{\chi}\frac{d\phi}{d\chi}+\beta\frac{d\phi}{d\chi}=1+\frac{(\phi^2-v^2)}{\chi^2}.\label{EXPLI}
\end{eqnarray} As it can be readly verified, Eq. (\ref{EXPLI}) can be written as
\begin{eqnarray}
\frac{d}{d\chi}\Bigg[\Bigg(\frac{\phi^2-v^2}{\chi}\Bigg)+\beta\phi\Bigg]=1,\nonumber
\end{eqnarray} whose integration gives
\begin{eqnarray}
\phi^2-v^2+\beta\phi\chi=\chi^2+c_0\chi,\nonumber
\end{eqnarray} where $c_0$ is an integration constant, which is the cause of the appearance of two-kink solutions and also responsible for a kind of split brane mechanism. Solving the above equation for $\chi$ as a function of $\phi$, we get
\begin{eqnarray}\label{orbit3}
 \chi(\phi) = \frac{\beta\phi + v b\sqrt{\beta^2 + 4} - f(\phi)}{2} ,
\end{eqnarray}
where the reparametrization $c_0 = - b\sqrt{\beta^2 + 4}$ was used, and we have defined
\begin{equation}
f(\phi) = \sqrt{(\phi \sqrt{\beta^2 + 4} + b v \beta)^2 + 4v^2(b^2 - 1)}. \nonumber
\end{equation}

In equation (\ref{orbit3}), the positive root has been disregarded, since it leads to non-localized energy density solutions. A careful analysis will reveal that the condition $b^2 > 1$ must be fulfilled to allow the existence of kink-like solutions. Substituting equations (\ref{superpotential}) and (\ref{orbit3}) into the first equation of (\ref{firstorder}), we have
\begin{equation}
\frac{d\phi}{dr} = \frac{1}{2}\{(\beta \phi + v b \sqrt{\beta^2 + 4}) f(\phi) - [f(\phi)]^2 \},
\end{equation}
The above equation can be solved analytically, and result reads
\begin{eqnarray}
 \phi(r) = v \sqrt{\frac{b^2 - 1}{\beta^2 + 4}} \bigg[ \frac{2 \tanh[v(r-r_0)] + 2 b}{\sqrt{b^2 - 1} (\sqrt{\beta^2 + 4} - \beta)} - \\ \nonumber - \frac{\sqrt{b^2 - 1} (\sqrt{\beta^2 + 4} - \beta)}{2 \tanh[v(r-r_0)] + 2 b} - \frac{\beta b}{\sqrt{b^2 - 1}} \bigg],
\end{eqnarray}
where $r_0$ is an integration constant associated with the translational invariance. By direct substitution of the last equation in (\ref{orbit3}), we obtain the explicit expression for $\chi(r)$
\begin{eqnarray}
\chi(r) = v \sqrt{\frac{b^2 - 1}{\beta^2 + 4}} \bigg[ \frac{\tanh[v(r-r_0)] + b}{\sqrt{b^2 - 1}} - \nonumber \\ - \frac{\sqrt{b^2 - 1}}{\tanh[v(r-r_0)] + b} + \frac{2 b}{\sqrt{b^2 - 1}} \bigg].
\end{eqnarray}
It is interesting to note that the behavior of the solutions $\phi(r)$ and $\chi(r)$ are strongly related to the integration constant $b$ and the parameter $\beta$. As one can see in Fig.2, the constant $b$ controls the two-kink behaviour of the solution $\phi(r)$ and it is also responsible to a flat-top region on the lump-like solution of the other field. Another interesting feature of the solutions occurs when $b$ assumes the critical value $b=1$. In such case both solutions $\phi(r)$ and $\chi(r)$ exhibit kink-like profile. On the other hand, in Fig. 2 we can see that the parameter $\beta$ breaks the symmetry of the solutions, and this feature will be very important in addressing the hierarchy problem.

\begin{figure}[H]
\center
\label{fig2}
\includegraphics[width=8cm]{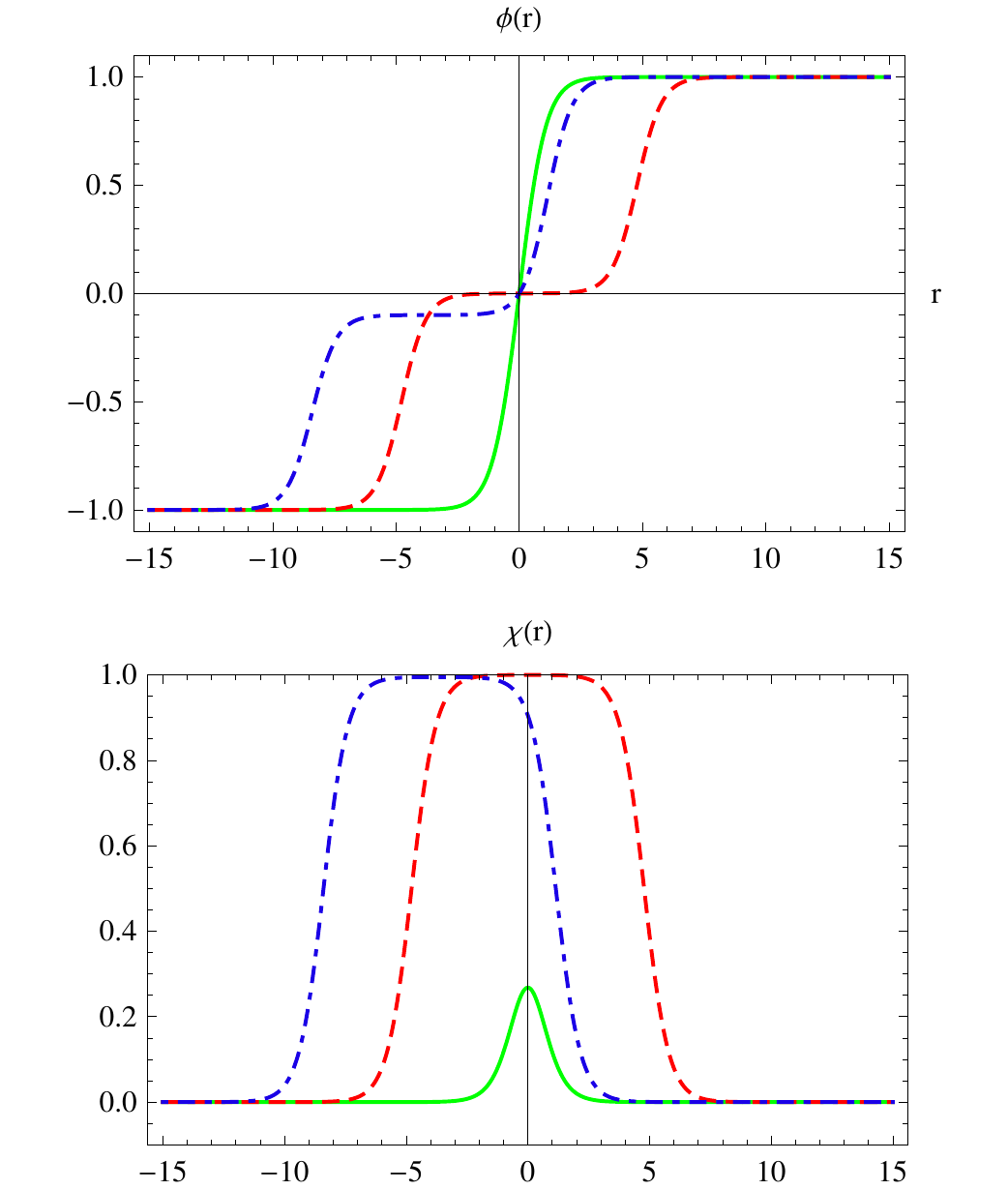}
\caption{Configurations of the fields $\phi(r)$ (above) and $\chi(r)$ (bellow) where $b = 2$, $v = 1$ and $\beta = 0$ (solid line); $b = 1.00000001$, $v = 1$ and $\beta = 0$ (dashed line); $b = 1.00000001$, $v = 1$ and $\beta = 0.2$ (dotted-dashed line).}
\end{figure}

According to our interpretation, the two-kink configuration corresponds to the formation of two 3-branes, each one localized at the same region of each kink. It is not difficult to conclude that the core of each brane is localized at the points in which $d\phi/dr$ takes its maximum value. Let us  designate $r_1$ and $r_2$ to be the core of each brane, at those points we must have $d^2 \phi/dr^2 = 0$. We note that a complete analytical expression of $r_1$ and $r_2$ in terms of the parameters $\beta$, $v$ and $b$ is not possible, but, we may obtain approximated expression with the following conditions
\begin{equation}
\phi(r_1) = \frac{v}{2}\frac{\sqrt{\beta^2 + 4} - \beta}{\sqrt{\beta^2 + 4}} \quad , \quad \phi(r_2) = -\frac{v}{2}\frac{\sqrt{\beta^2 + 4} + \beta}{\sqrt{\beta^2 + 4}}. \nonumber
\end{equation}
With the above equations, we get
\begin{equation}
r_i = r_0 + \frac{1}{v} \tanh^{-1}\bigg[ \frac{\sqrt{b^2 - 1}(\sqrt{\beta^2 + 4} - \beta)\Lambda_i - 2 b}{2}\bigg] , \,\nonumber
\end{equation}
where $i = 1,2$, and we have defined
\begin{equation}
\Lambda_i = \frac{\xi_i  + \sqrt{\xi_i^2 + 4 }}{2} \quad , \quad \xi_i = \frac{\sqrt{\beta^2 + 4} \phi(r_i) + v b \beta}{v \sqrt{b^2 - 1}} \nonumber .
\end{equation}
Despite the approximation character in the expression of $r_1$ and $r_2$, we may verify numerically that those expressions has an accuracy of $10^{-5}$ in its numerical value. Thus, any effect of this approximation in the scalar fields and in the warp factor are negligible.

In order to solve the last equation in (\ref{firstorder}) we perform the following manipulations
\begin{eqnarray}\nonumber
\frac{dA}{dr} &=& -\frac{2}{3} W(\phi,\chi) \\
              &=& -\frac{2}{3} \bigg[ v^2 \phi - \frac{1}{3} \phi^3 - \phi \chi^2 - \frac{\beta}{3} \chi^3 \bigg] \nonumber \\
              &=& -\frac{2}{9} [ \phi (v^2 - \phi^2 - \chi^2) + \chi (-2\phi\chi - \beta \chi^2) + 2 v^2 \phi ] \nonumber \\
              &=& -\frac{2}{9} [ \phi W_{\phi} + \chi W_{\chi} + 2 v^2 \phi] \nonumber \\
              &=& -\frac{2}{9} \bigg[ \phi \frac{d\phi}{dr} + \chi \frac{d\chi}{dr} + 2 v^2 \phi \bigg] \nonumber \\
              &=& -\frac{1}{9} \frac{d}{dr}(\phi^2 + \chi^2) - \frac{4 v^2 \phi}{9},\nonumber
\end{eqnarray}
thus, after integration we get
\begin{equation}
A(r) = -\frac{1}{9}[\phi(r)^2 + \chi(r)^2] - \frac{4 v^2}{9} \int dr \phi(r) + C ,
\end{equation}
where $C$ is an integration constant. By direct substitution of $\phi(r)$ and $\chi(r)$ on the above equation it is possible to obtain explicitly the analytical form of $A(r)$, but we will omit this expression here. Finally, the warp factor is given by $e^{2A(r)}$. We may fix the integration constant $C$ in order to ensure that $A(r_1) = 0$ and in this case we obtain $e^{2A(r_1)} = 1$. In Fig. 3 we plot the qualitative behavior of the warp factor. As we will see, the warp factor shape may be suitable for approaching the hierarchy problem.

\begin{figure}[H]
\center
\includegraphics[width=8cm]{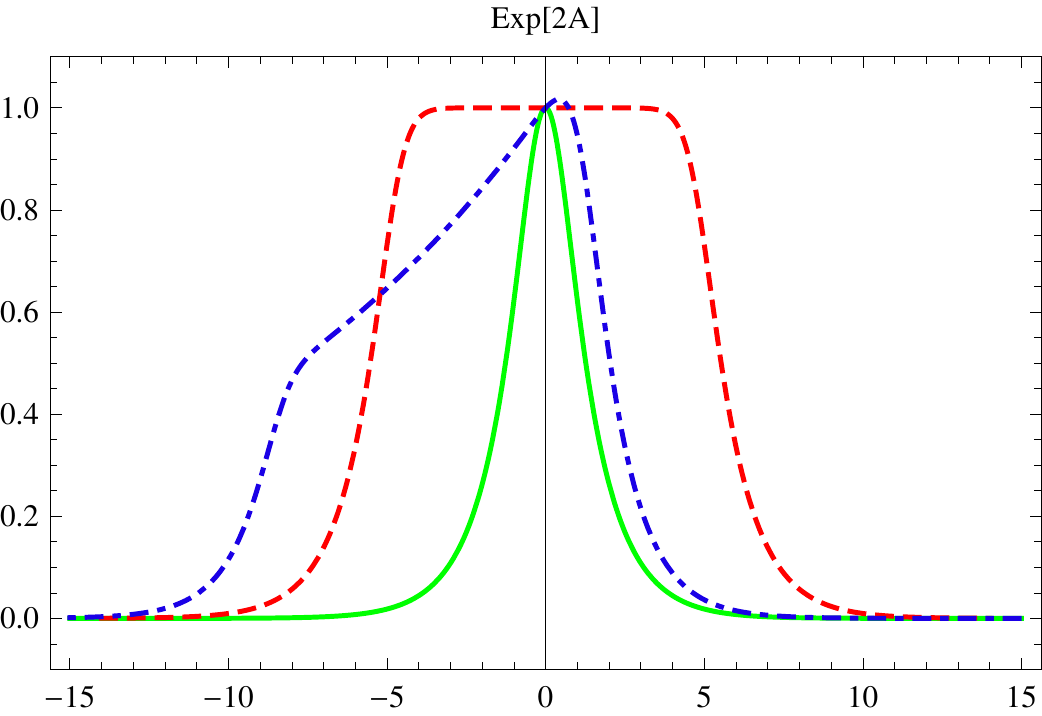}
\label{fig3}
\caption{Picture expliciting the warp factor qualitative behavior for the cases where $b = 2$, $v = 1$ and $\beta = 0$ (solid line); $b = 1.00000001$, $v = 1$ and $\beta = 0$ (dashed line); $b = 1.00000001$, $v = 1$ and $\beta = 0.2$ (dotted-dashed line).}
\end{figure}

In figures 4 and 5 we plot the graphics of the energy density and the scalar curvature, respectively. In the last one we may see how the parameters $\beta$, $v$ and $b$ affects the geometry of the bulk.

We note parenthetically that the energy associated with the scalar field is null. In fact, the expression for the energy density is given by
\begin{equation}
 \varepsilon(r) = e^{2A(r)}\bigg[ \frac{1}{2}\bigg( \frac{d\phi}{dr} \bigg)^2 + \frac{1}{2}\bigg( \frac{d\chi}{dr} \bigg)^2 + V(\phi,\chi) \bigg].
\end{equation}
Using the equations (\ref{firstorder}) and (\ref{pot}) we can rewrite the above equation as follows
\begin{equation}
 \varepsilon(r) = \frac{d}{dr}[e^{2A(r)}W(\phi,\chi)].
\end{equation}
Performing the integration with respect to the extra dimention, we get
\begin{equation}
\int_{-\infty}^{\infty} dr \varepsilon(r) = e^{2A(r)}W(\phi,\chi) \bigg|_{r=-\infty}^{r = \infty} = 0,
\end{equation}
since the warp factor vanishes when $r \to \pm \infty$. Therefore, the energy associeted with the scalar field is null.

\begin{figure}[H]
\center
\includegraphics[width=8cm]{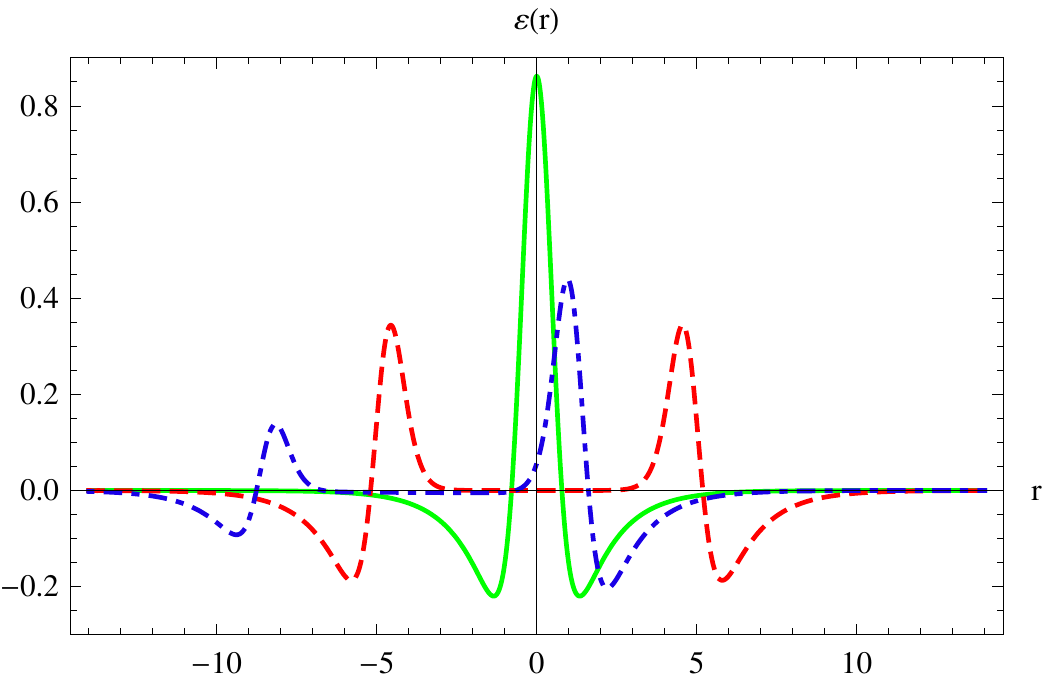}
\label{fig4}
\caption{Energy density for the cases $b = 2$, $v = 1$ and $\beta = 0$ (solid line); $b = 1.00000001$, $v = 1$ and $\beta = 0$ (dashed line); $b = 1.00000001$, $v = 1$ and $\beta = 0.2$ (dotted-dashed line).}
\end{figure}

\begin{figure}[H]
\center
\includegraphics[width=8cm]{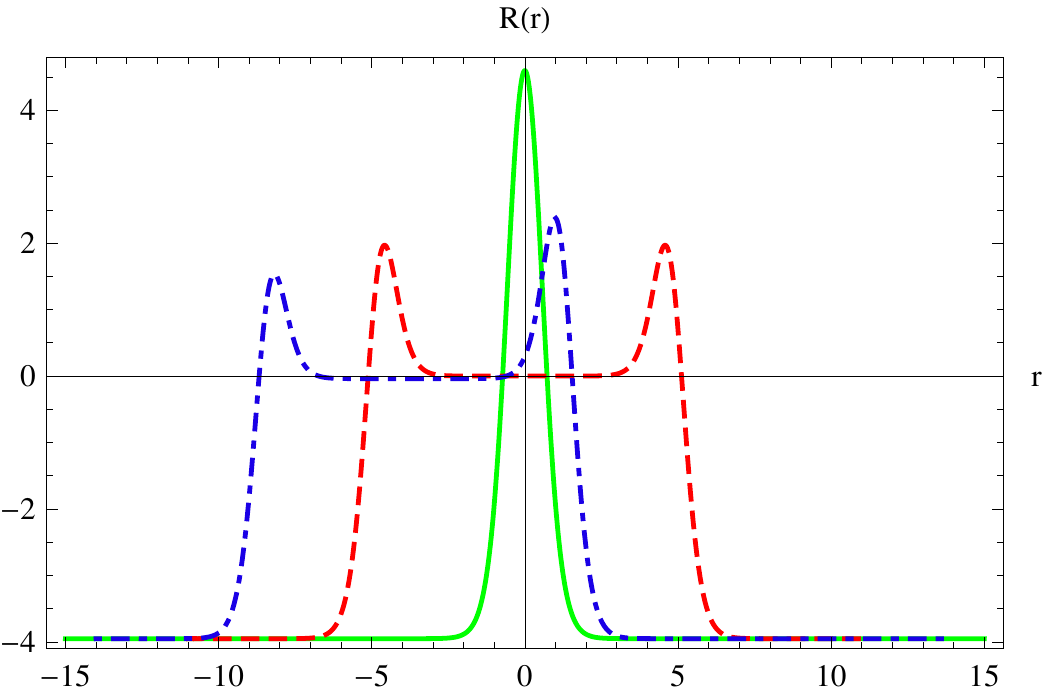}
\label{fig5}
\caption{Scalar curvature where $b = 2$, $v = 1$ and $\beta = 0$ (solid line); $b = 1.00000001$, $v = 1$ and $\beta = 0$ (dashed line); $b = 1.00000001$, $v = 1$ and $\beta = 0.2$ (dotted-dashed line).}
\end{figure}

Before going further, we would like to point out an important consequence of the asymmetric warp factor as the one shown in the dotted-dashed line of the Fig. 3. Generally speaking, it may be used to approach the hierarchy problem in the brane whose core is in $r<0$. Let us first make clear two assumptions concerning the usual picture. First, the hierarchy problem is usually analyzed in the context of thin (instead of thick) branes \cite{PAPERJHEP}. In our model, it is conceivable the introduction   of an additional parameter in the superpotential endowed with a specific limit in which the warp factor become more and more narrowed, leading to a thin brane-like setup. It is however not necessary. The point is that for each four-dimensional hipersurface ($r$ constant) we have a different value for the warp factor. In fact, the warp factor, when understood as a parameter entering in the particle masses via the Higgs mechanism, acts as a running parameter whose value (to be specified by the brane position
along the extra dimension) sets the energy scale experienced by brane observers. Hence, it is possible to investigate the consequences at the brane core, for instance. In this vein, all the possible modifications in the observables due to the brane thickness, if any, shall be constrained by the experiment. Another remarkable point is that we shall assume the localization of the Higgs scalar field on the brane. The localization issue is out of this paper's scope, nevertheless it is well known that the localization of any scalar field can be achieved by means of the warp factor, which furnishes the necessary gravitational weight to localize the field.

Having said that, the analysis of the hierarchy problem is quite usual \cite{RS1,HER,PAPERJHEP}. We shall make explicit its general steps here, however, for book keeping purposes. Let $\Phi$ be the Higgs field. The action describing the simple Higgs at the brane core is given by
\begin{equation}
S\!\!\sim\!\!\!\! \int\!\!\! d^4x\sqrt{-G} \{G^{\mu\nu}\partial_{\mu}\Phi^{\dagger}\partial_{\nu}\Phi-m^2\Phi^{\dagger}\Phi+\lambda (\Phi^{\dagger}\Phi)^2\}.\label{hp1}
\end{equation} If the brane in question is the one whose core is in $r=r_1$, then everything goes as in the Minkowski space. However, for the brane placed at, say, $r= r_2 = \bar{r}$ we have
\begin{eqnarray}
S\!\!\sim\!\!\!\int\!\! d^4x e^{4A(\bar{r})}\{e^{-2A(\bar{r})}\partial^\mu\Phi^\dagger\partial_{\mu}\Phi-m^2\Phi^{\dagger}\Phi+\lambda (\Phi^{\dagger}\Phi)^2\}.\nonumber
\end{eqnarray} Hence, after the rescaling the Higgs field\footnote{This rescaling is always possible. Nevertheless it is necessary to investigate the physical consequences it brings by looking at the observables.} as $\Phi\rightarrow e^{-A(\bar{r})}\Phi$ one has
\begin{equation}
S\sim \int d^4x\{\partial^\mu\Phi^\dagger\partial_{\mu}\Phi-\bar{m}^2\Phi^{\dagger}\Phi+\lambda (\Phi^{\dagger}\Phi)^2\},\label{hp2}
\end{equation} where $\bar{m}^2=m^2e^{2A(\bar{r})}$ is the revested parameter entering in the masses distribution. The result is, then, exhaustive: any mass parameter measured by a $r=\bar{r}$ core brane observer will be revested by the factor $e^{2A(\bar{r})}$. Since the warp factor value has a damping at this region (see Fig. 3), it is possible to associate the warp factor damping to the gap between Planck and TeV scales. In other words, if the observer on the $r=r_1$ brane core experiences Planck scales, the observer on the $r=\bar{r}$ brane core experiences a damped energy scale, whose value can be easily adjusted by the model parameters to be of TeV order.  We emphasize once again that the effects of the brane thickness on the observables, just as the masses, shall be constrained by high energy experiments --- in the electroweak sector, for instance --- making its possible departures from the above analysis quite small.

Finally, in Fig. 6 we depict some special ranges in the $\beta \times v$ parameter space where the warp factor is suitable for address the hierarchy problem, that is $e^{A(\bar{r})}\sim 10^{-15}$.  The situation, therefore, goes as follows: concerning the warp factor behavior, the integration constant $b$ controls the width of the flat-top, while the $\beta$ parameter induces an asymmetrical shape. Hence, as the explicited ranges of Fig. 6 indicate, we have dense regions where the warp factor can be used in order to connect Planck and TeV scales, precluding the necessity of fine-tuning in any sense.

\begin{figure}[H]
\center
\includegraphics[width=7cm]{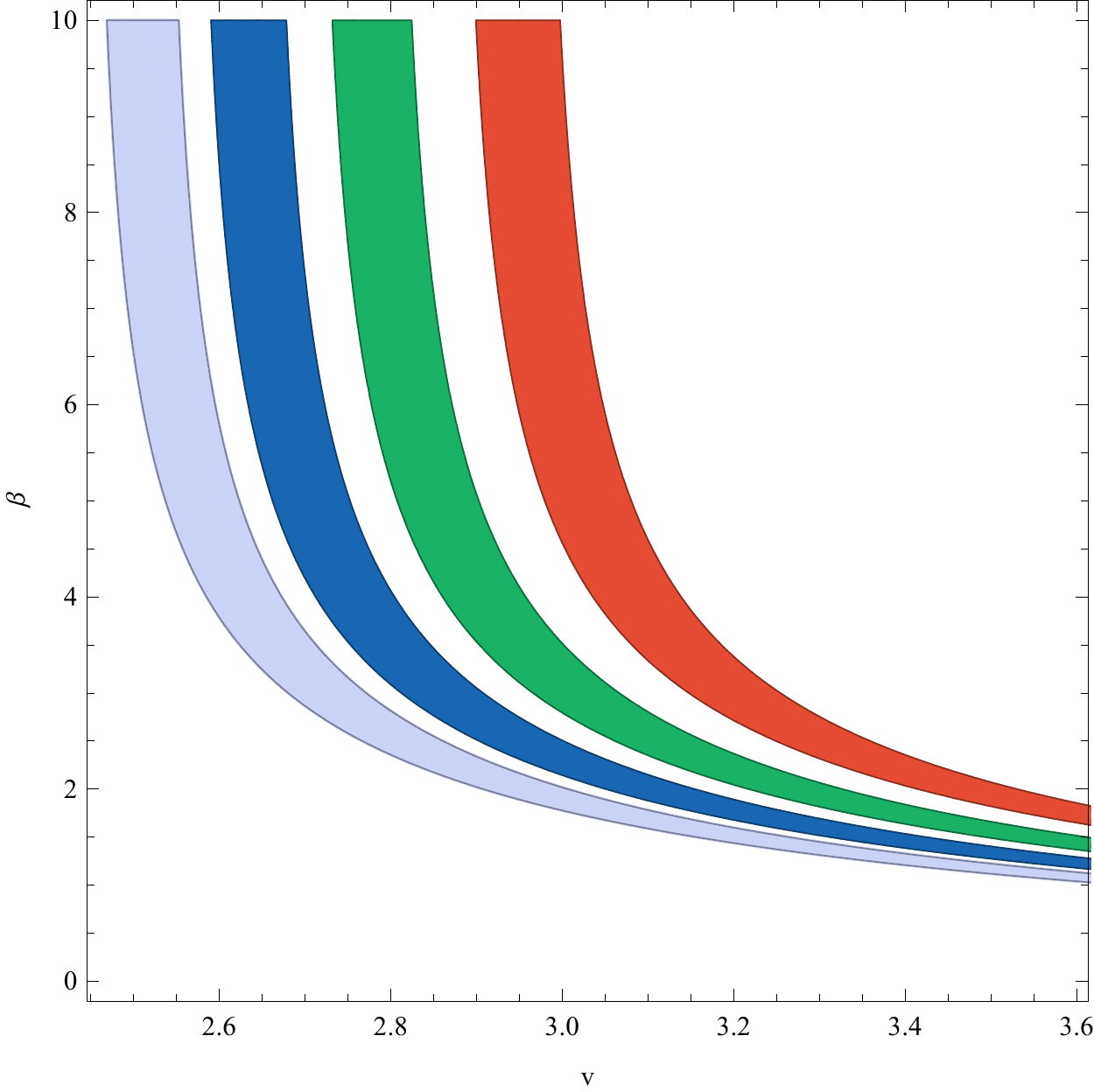}
\label{Hregion.pdf}
\caption{Important ranges in the $\beta \times v$ parameter space, within which the warp factor is suitable to approach the hierarchy problem ($e^{A(\bar{r})}\sim 10^{-15}$). Each region corresponds to a specific value of the parameter $b$. From left to right $b = 1,00000000001$; $b = 1,0000000001$; $b = 1,000000001$; $b = 1,00000001$.}
\end{figure}

\section{Stability and metric fluctuation}

In this section we investigate the metric fluctuation including fluctuations around the scalar field background presented in the previous section. For the metric fluctuation we adopt a gauge where the perturbed interval becomes \cite{GREMM,DEWOLFE}
\begin{equation}
 ds^2 = e^{2A(r)}(\eta_{\mu\nu} + h_{\mu\nu})dx^{\mu}dx^{\nu} + dr^2,
\end{equation}
with $h_{\mu\nu} = h_{\mu\nu}(r,x_{\alpha})$. We are concerned with small metric fluctuations. Keeping only terms in the action up to second order we obtain a complicated system of differential equations. However, there is a sector, the transverse and traceless $h_{\mu\nu}$, in which the system of equations can be simplified. It can be done by using the projector operator
\begin{equation}
P_{\mu \nu \alpha \beta }\equiv \frac{1}{2}\left( \pi _{\mu \alpha }~\pi_{\nu \beta }+\pi _{\mu \beta }~\pi _{\nu \alpha }\right) -\frac{1}{3}\pi_{\mu \nu }~\pi _{\alpha \beta },
\end{equation}
where $\pi _{\mu \nu }\equiv \eta _{\mu \nu }-\frac{\partial _{\mu }\partial_{\nu }}{\square }$. In other words, by using $\bar{h}_{\mu \nu }= P_{\mu \nu \alpha \beta }~h^{\alpha \beta }$, we get
\begin{equation}
\frac{d^{2}\bar{h}_{\mu \nu }}{dr^{2}}+4\,\frac{dA}{dr}\,\frac{d\bar{h}_{\mu\nu }}{dr}-e^{-2\,A}\partial _{\rho }\partial ^{\rho}\bar{h}_{\mu \nu }=0.
\end{equation}
Performing a function redefinition
\begin{equation}
 \bar{h}_{\mu \nu }\equiv e^{i\,\vec{k}.\vec{x}}\,e^{-\frac{3}{2}\,A}\psi
_{\mu \nu },
\end{equation}
and a variable transformation $z=\int e^{-A\left( r\right) }dr$, we can recast the above equation as a quantum mechanics-like problem
\begin{equation}
-\frac{d^{2}\psi _{\mu \nu }}{dz^{2}}+U_{eff}\left( z\right) \psi _{\mu \nu}=k^{2}\psi _{\mu \nu },\label{ut}
\end{equation}
where the effective potential is defined by
\begin{equation}
U_{eff}\left( z\right) =\frac{9}{4}\left( \frac{dA}{dz}\right) ^{2}+\frac{3}{%
2}\frac{d^{2}A}{dz^{2}}.\label{il}
\end{equation}
In terms of the old variable, $U_{eff}$ can be written as
\begin{equation}
U_{eff}(r) =\frac{3}{4}e^{2A}\left( 2\frac{d^{2}A}{dr^{2}} +5 \left( \frac{dA}{dr}\right) ^{2}\right).
\end{equation}
For completeness, in Fig. 7 we plot $U_{eff}(r)$ for three different values of the parameters $\beta$ and $b$.

\begin{figure}[H]
\center
\includegraphics[width=8cm]{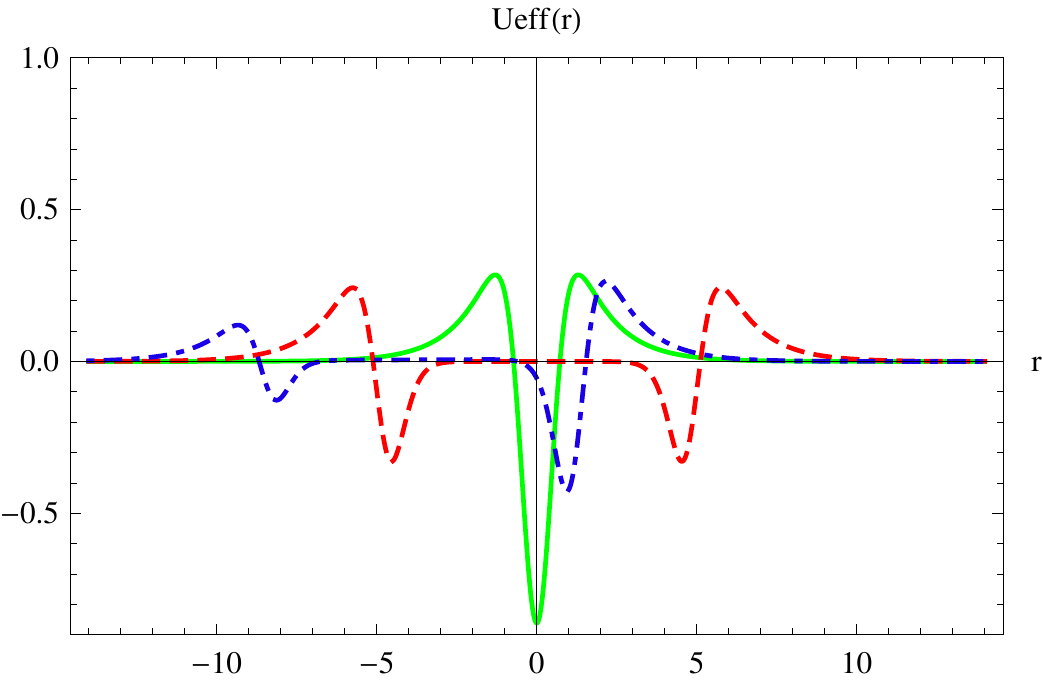}
\label{fig6}
\caption{Effective potential $U_{eff}(r)$ with $b = 2$, $v = 1$ and $\beta = 0$ (solid line); $b = 1.00000001$, $v = 1$ and $\beta = 0$ (dashed line); $b = 1.00000001$, $v = 1$ and $\beta = 0.2$ (dotted-dashed line).}
\end{figure}

Following \cite{PAPERJHEP} it is easy to note that Eq. (\ref{ut}), along with Eq. (\ref{il}), may be recast into the form
\begin{equation}
\mathbb{O}^\dagger\mathbb{O}\psi_{\mu\nu}=k^2\psi_{\mu\nu},\label{SS}
\end{equation} where $\mathbb{O}^\dagger\mathbb{O}=\Bigg(-\frac{d}{dz}-\frac{3}{2}\frac{dA}{dz}\Bigg)\Bigg(\frac{d}{dz}-\frac{3}{2}\frac{dA}{dz}\Bigg)$. Therefore, since $\mathbb{O}^\dagger\mathbb{O}$ is positive, the supersymetric version of (\ref{SS}) ensure the absence of tachyonic modes.

\section{Concluding Remarks}

We presented a model consisting of two scalar fields, coupled to gravity, whose dynamics lead to a well established thick braneworld model. We investigate the dynamical system behind the first order equations for arbitrary parameters. It was shown that there are non trivial fixed points acting as attractor points for some positive values of $\lambda$, taking into account the right region for the $\beta$ parameter (see Fig. 1).

It was shown that for a particular case in the parameters space, namely $\lambda=1$, it is possible to find out an analytic solution for the coupled system, by means of the so-called orbit equation procedure. In this case, we arrive at a resulting asymmetric warp factor (Fig. 3) which can be used to approach the hierarchy problem. We stress that there is a dense set of points in the parameter space in which the warp factor is adequate to explain the hierarchy problem without fine-tuning (Fig. 6).

By investigating the metric fluctuation, the absence of negative mass modes under small perturbations, ensures the model stability. In general grounds, we presented a consistent braneworld model performed by two scalar fields obeying a nontrivial (super)potential constraint. The resulting model, in a specific range of the parameters, allows for approaching of the hierarchy problem.

\section*{Acknowledgments}
The authors are grateful to CNPq and Fapesp for the financial support, and thanks to prof. M. Hott for enlightening discussions concerning braneworlds scenarios.

\end{document}